\documentclass[lettersize,journal]{IEEEtran}
\usepackage{amsmath,amsfonts}
\usepackage{algorithmic}
\usepackage{algorithm}
\usepackage{array}
\usepackage[caption=false,font=normalsize,labelfont=sf,textfont=sf]{subfig}
\usepackage{textcomp}
\usepackage{stfloats}
\usepackage{url}
\usepackage{verbatim}
\usepackage{graphicx}
\usepackage{cite}
\usepackage{subcaption}
\usepackage{hyperref}
\usepackage{graphicx}
\usepackage{subcaption}
\usepackage{multirow}
\usepackage{tabularx}
\usepackage{array}
\usepackage{float}
\usepackage[T1]{fontenc}

\begin{document}

\title{Modeling Learner–Technology Interaction Through Multilevel Gaze Transition Dynamics to Inform Usability in Online Learning}

\author{Yueqi Weng, Jie "Jennifer" Lu, Fan Yang
\thanks {Yueqi Weng and Jie "Jennifer" Lu are with the University of Georgia, Georgia,
United States}
\thanks{Fan Yang is with the University of Arizona, Arizona, United States}
}
\maketitle

\begin{abstract}
Gaze data provide fine-grained evidence of how learners engage with instructional content and interface elements, complementing conventional usability evaluation methods. However, eye-tracking analyses often examine within-slide visual attention and across-slide navigation separately, limiting understanding of learner interaction throughout extended learning experiences. This exploratory multi-method study integrated area-of-interest transitions, gaze entropy, slide-level navigation, and network-based representations to analyze seven eye-tracking recordings of high school students interacting with an AI-supported online biology lesson, complemented by three post-intervention focus groups. Results showed individual differences in within-slide gaze allocation and switching patterns. Learners followed shared navigation pathways but differed in revisiting and immediate return behaviors. The focus group findings highlighted positive perceptions of interactive materials and AI-generated feedback, alongside usability concerns related to navigation, interface clarity, and assessment guidance. Together, these outcomes show how integrating multilevel gaze analytics with learner perspectives can provide complementary evidence for identifying interaction trends relevant to usability evaluation. Future research should test this approach with larger, more diverse samples and examine relationships among gaze patterns, usability, and learning outcomes.
\end{abstract}


\section{Introduction}
\IEEEPARstart{O}{nline} learning environments can reflect continuous learners' behavioral trace data as they interact with instructional content, yet these interactions are often represented through limited pathways, for example, observable indicators such as clicks, page visits, and task duration \cite{bergdahl2024}. Such indicators provide evidence of learner activity but capture only part of how learners engage with a learning intervention\cite{winter2024}. Learning engagement is multidimensional and context-dependent \cite{reeve2025,wong2024}, so behavioral indicators need to be interpreted in relation to the learning activity and its surrounding context \cite{demmler2026}. Eye tracking provides a complementary source of evidence by recording the spatial and temporal allocation of visual attention \cite{liu2025}. By preserving information about where attention is directed and how it changes over time, gaze data provide a process-oriented representation that unfolds learners' online interaction and complements action-based behavioral trace measures \cite{saizmanzanares2025,kunz2024}.

This process-oriented perspective makes gaze relevant to the study of learning experience and usability in digital environments\cite{molina2025,novak2024}. Eye movements reflect the ongoing allocation of visual attention during human--computer interaction and have long been used to examine how users perceive and interact with interface elements \cite{majaranta2014}. In educational contexts, eye tracking has been used to investigate cognitive processing and engagement with complex visual stimuli \cite{liu2025}, moment-to-moment attention during interaction with instructional content, and cognitive load in multimedia learning \cite{shahnabati2025}. Consequently, gaze data can complement conventional behavioral traces by providing evidence on visual processes through which learners encounter, inspect, and navigate instructional information\cite{stark2024,gu2024,tothova2025}. For usability evaluation, these data can help characterize how learners visually interact with instructional materials and interface elements, providing behavioral evidence that complements their reported experiences.

Realizing this potential becomes difficult when eye-tracking analysis shifts from isolated tasks toward continuous and dynamic learning sessions \cite{larigaldie2024,blascheck2017,ke2024,schindler2025}. Common visual analysis methods such as heatmaps and scanpaths are useful for characterizing specific tasks and intervals, but these visualizations are challenging to capture the multiple dimensions of eye-movement behavior at once \cite{claus2025}. When applied across an extended online intervention, analyses conducted separately for individual activities, area of interest (AOIs), or time intervals can fragment temporally structured gaze behavior, making it difficult to trace how visual attention develops across the interaction as a whole\cite{fu2024,vojtechovska2026}. Manual review and annotation of large volumes of gaze data are also labor-intensive \cite{le2025}, limiting the identification of recurring structures across extended sessions without computational support \cite{selim2024}. Thus, the analytical challenge mainly lies in how to transform continuous gaze traces into representations that retain meaningful patterns of learner interaction across time.

Another limitation concerns the reliance on aggregate gaze measures. Eye-tracking research commonly summarizes attention using fixation count, fixation duration, and dwell time to predefined AOIs \cite{liu2025,hou2025,suzuki2024}. These measures connect gaze behavior to certain design elements by quantifying where attention is allocated and how long it remains there \cite{tothova2025}. However, aggregation within AOIs provides limited information about the temporal order in which attention moves between elements \cite{cui2024,niehorster2025}. Temporal ordering is important because transitions among AOIs characterize how visual attention shifts, returns, and develops as learner interaction unfolds \cite{niehorster2025,laborde2026}. Learners can exhibit comparable aggregate attention to the same interface elements while differing substantially in the sequences through which those elements are viewed\cite{laborde2026}. This way, aggregate gaze measures alone may obscure differences in the temporal organization and dynamics of learners' visual attention.

Sequence-sensitive measures are promising for addressing the temporal organization of gaze. Recent research \cite{han2026extending} combined conventional AOI measures with entropy- and velocity-based metrics to characterize learner attention in an online environment. Gaze transition entropy quantifies the predictability of gaze transitions among AOIs \cite{krejtz2015,cui2024}. Scanpath-based approaches further represent gaze as an ordered sequence, enabling analysis of transitions, recurrence, and similarity across viewing patterns \cite{laborde2026}. However, a remaining challenge is linking gaze behavior across two nested levels: shifts of visual attention among elements within a page of an online program, and navigation across pages over the whole learning session \cite{fan2023,huber2024}. Within pages, gaze moves among instructional and interface elements, while across pages, navigation reflects how learners move through and revisit the broader environment. Accordingly, AOI analyses capture local visual behavior but not broader navigation trajectories, whereas page-level traces capture navigation without representing within-page gaze dynamics. Existing approaches to link these levels of gaze analytics into a continuous representation of learner online interaction remain limited, especially for examining usability across an extended digital learning experience\cite{dever2023,tao2025}.

This study addresses this gap by modeling gaze as a time-varying state within user interface design elements and examining learner interaction through multilevel gaze dynamics. The proposed framework integrates AOI, gaze transition, recurrence, entropy-based metrics, and temporal measures to connect within-page visual behavior with between-page navigation. We apply this framework to an AI-supported online lesson used by high school students in a research hospital education program and triangulate gaze patterns with post-intervention individual debrief data to interpret their usability implications in relation to learners' reported experiences.

\section{Literature Review}

\subsection{Usability Evaluation of Online Learning Environments}
Usability is commonly defined, following the international standard ISO 9241-11, as ``the extent to which a product can be used by specified users to achieve specified goals with effectiveness, efficiency and satisfaction in a specified context of use'' \cite[p.~536]{bevan2001international}. In online learning environments, users are also learners \cite{zaharias2009developing}, so functional ease of use alone does not show whether an environment supports learning. Usability matters because the usability of a technology interface may have a major influence on learning \cite{davids2014usability}, and perceived usability affects both learners' experiences and their learning outcomes \cite{vlachogianni2022perceived}. When a system is difficult to use, learners may spend more time learning the system than the content \cite{ardito2006approach}. This concern is especially prominent in online learning environments, where technology serves as the primary medium of instruction. A usability evaluation typically specifies what is assessed and how. Regarding what is assessed, evaluations commonly address effectiveness, efficiency, and learner satisfaction \cite{abuhlfaia2018usability,vlachogianni2022perceived}. Educational researchers have extended these technological criteria to pedagogical aspects, such as content, instructional strategy, and learner support, and socio-cultural aspects, such as communication and collaboration \cite{lu2022usability}. For e-learning specifically, Ardito et al.~\cite{ardito2006approach} argued that evaluation must jointly consider usability, accessibility, and didactic effectiveness. Zaharias and Poylymenakou~\cite{zaharias2009developing} moved beyond functional usability by combining web and instructional design parameters, such as navigation, visual design, and learner guidance and support, with intrinsic motivation to learn as an affective usability measure. Regarding how usability is assessed, evaluations draw on three main classes of methods \cite{lu2022usability}. Inquiry methods, such as questionnaires and interviews, ask users to report their experiences; inspection methods, such as heuristic evaluation, have evaluators examine an interface against established criteria; and testing methods, such as task-based user testing and think-aloud protocols, involve observing users as they complete tasks. These evaluations span e-learning, immersive, mobile, and game-based technologies, with e-learning accounting for nearly two-thirds of studies \cite{lu2022usability} and higher education being the main context in which Internet-based learning platforms were evaluated \cite{vlachogianni2022perceived}.

Despite this growing body of work, current usability research on learning technologies shows several methodological drawbacks. First, it relies heavily on self-report. Inquiry methods accounted for 64.8\% of methods in the review by Lu et al.~\cite{lu2022usability}, and questionnaires alone made up nearly half. Studies of e-learning platforms in higher education likewise relied mostly on simple research designs dependent on questionnaires, which were used in 50 of the 61 studies reviewed \cite{abuhlfaia2018usability}. However, questionnaires capture perceptions of use rather than actual use \cite{lu2022usability}, and self-reports can mislead. For example, Davids et al.~\cite{davids2014usability} found an e-learning resource unusable for many participants, despite good satisfaction ratings reported in an earlier evaluation. Second, methodological rigor is limited. In the sample reviewed by Lu et al.~\cite{lu2022usability}, many questionnaires and all task-based testing instruments were self-developed, and ``general usability'' was the most frequently evaluated technological outcome, typically without any explanation of which usability features were measured. Moreover, 77.9\% of studies relied on a single method, even though no single method captures all usability problems \cite{lu2022usability}. Third, outcome measures remain narrow. Technological usability accounted for over 85\% of measures, compared with 13.4\% for pedagogical and 1.4\% for socio-cultural usability \cite{lu2022usability}. Finally, evaluations are often conducted summatively, late in development, as confirmatory checks rather than as formative investigations of learner behavior \cite{lu2022usability}. These drawbacks leave limited evidence on how learners actually allocate attention and navigate content and interface elements during interaction. Process-oriented measures, which capture learner behavior during interaction, can address this gap. The E-TRACKING project, for example, evaluated e-learning systems through learners' exploratory eye movements, offering an objective and quantitative complement to mainly subjective and qualitative methods \cite{zambarbieri2006design}. In another study, Liu and Zhu~\cite{liu2012case} combined eye tracking with post-test interviews to identify which elements of an e-learning course homepage attracted or escaped attention, and found that the combination yielded a richer interpretation than either method alone. Building on such evidence, Lu et al.~\cite{lu2022usability} recommended eye-tracking metrics, such as fixation sequences and areas of interest (AOIs), noting that they can reveal non-conscious processes and design issues that traditional testing may miss.

\subsection{Eye Tracking and Gaze Analytics for Usability Evaluation}
Eye tracking has become increasingly common in educational technology research, particularly since 2014 \cite{lai2025trends}. For usability evaluation, eye tracking records where users look and the order in which their gaze moves across a display, without relying on self-report, thus offering a more objective source of evidence. Using such evidence to evaluate displays dates to a 1950 study by Fitts et al. of military pilots' eye movements during instrument-landing approaches, which proposed that fixation frequency on a display element reflects its importance, fixation duration the difficulty of extracting information from it, and transitions between elements the efficiency of their arrangement \cite{jacob2003eye}. These interpretations rest on what Just and Carpenter~\cite{just1980theory} later formalized as the eye--mind assumption: the eye remains fixated on an item as long as that item is being processed, so gaze reveals what a user is attending to and processing. Applying this assumption, usability researchers define AOIs over interface elements and analyze the eye movements within them to assess the visibility, meaningfulness, and placement of each element \cite{poole2006eye}. For example, an unusually long gaze on an icon before it is selected suggests that the icon lacks meaningfulness and may need redesign \cite{poole2006eye}. Such measures can distinguish design quality. Goldberg and Kotval~\cite{goldberg1999computer} found that a poorly organized interface produced longer scanpaths, more fixations, and denser transition matrices between AOIs than a well-organized one. They argued that eye-movement analysis can enhance traditional usability evaluations, which are often qualitative and of limited reliability. Reviewing five decades of such work, Jacob and Karn~\cite{jacob2003eye} concluded that eye tracking can contribute substantially to usability evaluation, though not as a complete approach on its own.

Early eye-movement-based usability studies relied heavily on cumulative fixation time within AOIs, and few examined the sequential structure of scanpaths \cite{goldberg1999computer}. Aggregate representations such as heatmaps likewise show which regions were fixated but not the order in which they were visited \cite{krejtz2015}. Yet order matters for usability. Two scanpaths with identical spatial coverage can have very different transition patterns: one direct and efficient, the other moving back and forth between areas in a way that indicates uncertainty \cite{poole2006eye}. Transition-based measures capture this sequential information. Goldberg and Kotval~\cite{goldberg1999computer} used transition matrices to represent eye movements between AOIs, treating a dense matrix as a sign of extensive search and a sparse matrix as a sign of more directed search. Krejtz et al.~\cite{krejtz2015} extended this approach by modeling each individual's AOI sequence as a first-order Markov chain and computing two normalized entropy measures ranging from 0 to 1. Transition entropy captures the predictability of switching between AOIs, and stationary entropy captures how evenly attention is distributed across AOIs. In multimedia learning, interactive and animated visualizations produced lower transition entropy than static illustrations, which the authors interpreted as more systematic, less chaotic inspection of the learning material \cite{krejtz2016gaze}.

Interpreting these measures, however, requires caution. The same metric can carry different meanings depending on the task. For example, a high fixation frequency on an area during browsing may indicate either interest in the content or difficulty understanding it. During searching, the same frequent fixations may instead reflect uncertainty about whether the area is the target the user is looking for \cite{poole2006eye}. Entropy measures are similarly context-dependent. In the art-viewing studies of Krejtz et al.~\cite{krejtz2015}, low transition entropy combined with high stationary entropy was tentatively linked to greater curiosity, whereas high values of both accompanied recognition of a familiar painting. Moreover, the difference in transition entropy between viewers who did and did not recognize the painting was significant for only one of six AOI grid sizes, suggesting that whether entropy analyses detect an effect can depend on how AOIs are defined. Interpretation becomes even more difficult in dynamic interfaces, where scrolling, pop-up messages, and user-initiated navigation make it harder to determine what users were looking at \cite{jacob2003eye}. In online learning environments, where learners move across many pages or slides, gaze patterns therefore need to be interpreted alongside learners' navigation.

\section{Method}
This exploratory multi-method research combined eye-tracking analytics with post-lesson focus groups to examine how learners interacted with an AI-supported digital lesson and to identify usability issues. Eye-tracking captured gaze allocation among interface elements and navigation across lesson slides. Focus group interviews were held immediately after learners explored the lesson, capturing their reported experiences and perceptions of usability. The research questions are stated below:
\begin{itemize}
    \item \textbf{RQ1:} How do learner within-slide behaviors differ from mul-
tilevel gaze data?
    \item \textbf{RQ2:} What navigation trajectories do learners follow through the digital learning intervention?
    \item \textbf{RQ3:} How can gaze patterns inform understanding of the usability of the digital learning environment?
\end{itemize}

\subsection{Learning Intervention Description}
  The learning environment is an online, case-based biology lesson that follows the treatment journey of a young patient with acute lymphoblastic leukemia (ALL). The lesson introduced cell-cycle regulation through instructional content and multiple interactive elements, including checkpoints, drag-and-drop activities, a video, and multiple-choice assessments, to help students connect and assess their understanding of cancer and treatment. GPT-4o powered an evaluation agent for this online intervention, providing personalized feedback during three open-response activities, including case-based reflection and lesson review. Guided by expert-authored reference answers, the feedback provides accurate explanations and areas for learners to improve their reflection. These activities provide learners with opportunities to explain biological concepts, apply their understanding to the patient case, and synthesize content across the lesson.

\subsection{Participants}
We enrolled 18 high school students aged 16 to 18 in an eight-week program at a research hospital. The program offers hands-on mentored research to rising high school seniors across biomedical disciplines. Participants had demonstrated interest in biomedical research and related educational or career pathways. We obtained informed consent from all participants and, for those under 18, from a legal guardian. After the eye-tracking sessions, we conducted three focus group interviews with participating students (\textit{n}=7) to ask about their experience and provide feedback on the lesson.

\subsection{Data Collection}

\subsubsection{Eye-Tracking Sessions}
Each participant independently explored the lesson for approximately 10--20 minutes on a laptop equipped with a Tobii Pro Spark eye tracker. The software exported individuals' recordings with gaze points and timestamps.
\subsubsection{Focus Group}
The research team facilitated post-intervention focus group interviews using a semi-structured format, covering overall experience, interaction and navigation, perceptions of the AI features, and suggestions for improvement. We recorded all focus group activities via Zoom and exported and labeled the recordings by participant order. 

\subsection{Data Analysis}

\begin{figure}[t]
    \centering
    \includegraphics[width=0.8\linewidth]{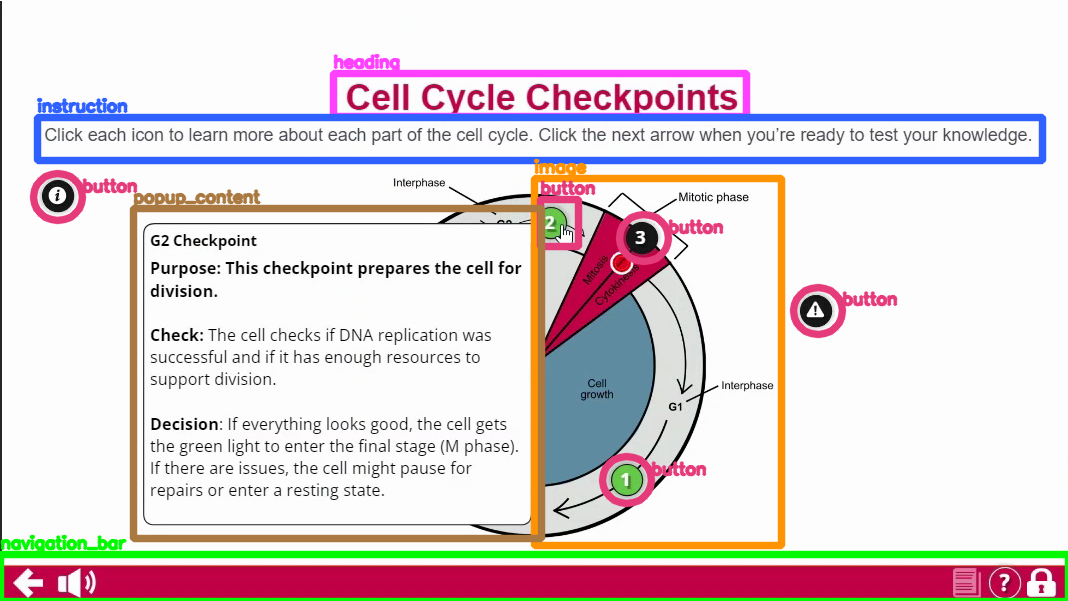}\\[4pt]
    \includegraphics[width=0.8\linewidth]{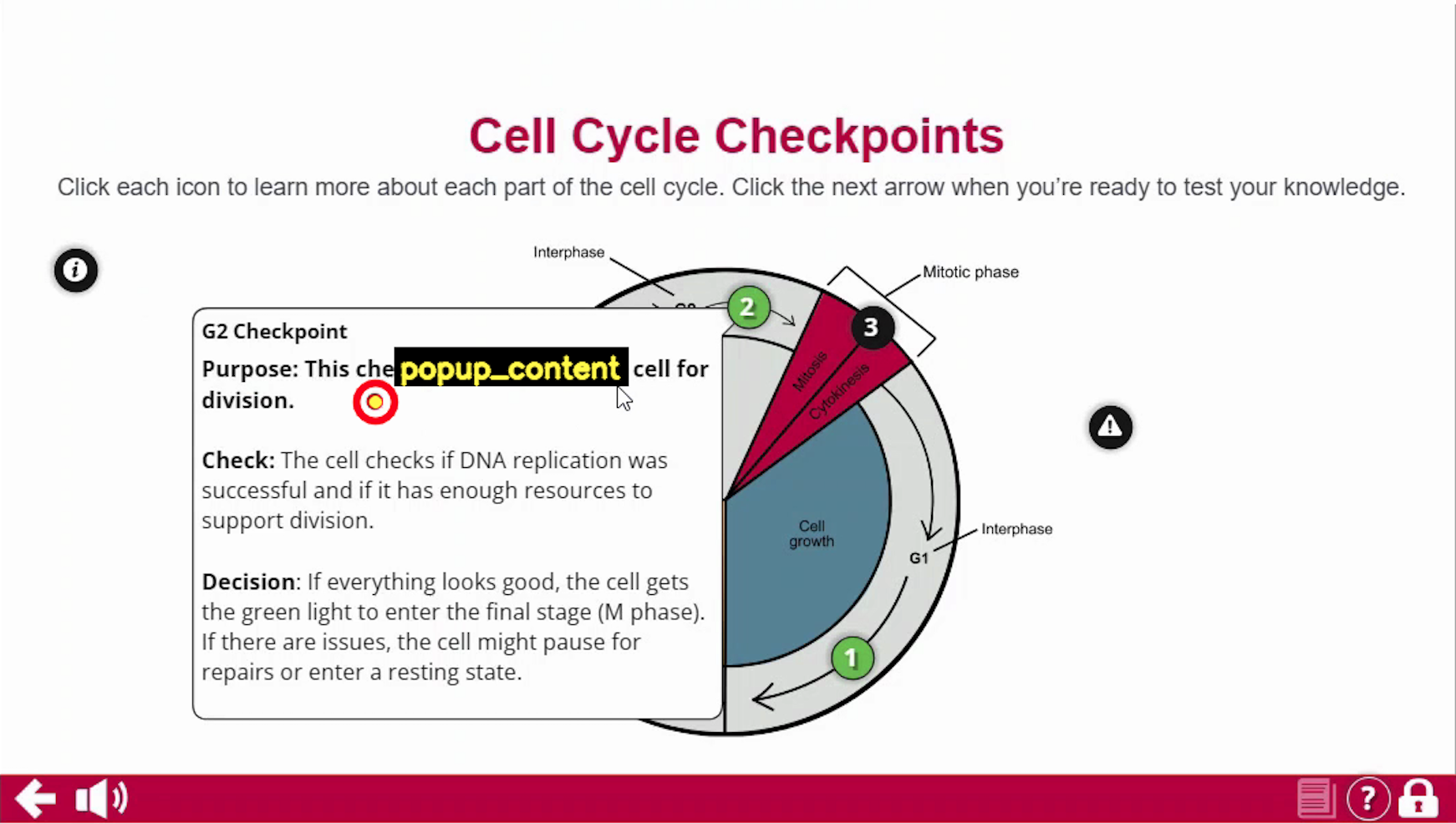}
    \caption{Predefined AOIs in the digital learning interface(top) and automated gaze recognition reference(bottom)}
    \label{fig:aoi_interfaces}
\end{figure}

We implemented the eye-tracking analysis pipeline in Python using OpenCV and NumPy for gaze detection, image matching, coordinate processing, and geometric analysis. A shared slide-element library was constructed once from a reference recording of the course and then reused across all participant recordings. We stored crops of visually stable course screens as slide references and assigned each screen a stable slide identifier, $S_m$. We labeled visually meaningful instructional and interface components, such as instructions, images, assessment regions, buttons, and navigation controls, as AOIs shown in Figure~\ref{fig:aoi_interfaces}. AOI boundaries were initially proposed using EasyOCR and OpenCV and subsequently reviewed and corrected by a researcher. Each AOI $E_{mn}$ was represented as a two-dimensional polygon in screen-image coordinates. The slide-element library defined the spatial structure of each slide,
\begin{equation}
S_m=\{E_0,E_{m1},E_{m2},\ldots,E_{mN_m}\},
\end{equation}
where $E_0$ denotes the slide-specific unrelated state and $N_m$
denotes the number of explicitly defined instructional and interface
AOIs on slide $m$.

Gaze position was first detected frame by frame from the red circular gaze indicator displayed in the screen recording. OpenCV-based color segmentation and contour analysis were used to estimate the gaze coordinates
\begin{equation}
G(t)=(x_t,y_t),
\end{equation}
where $x_t$ and $y_t$ denote the gaze position in screen-image coordinates at time $t$. The temporary failure to detect the gaze marker was treated as a detector miss. The missing coordinates were filled using a temporal hold procedure. We filled internal gaps from the most recent detectable gaze location, while assigning leading gaps the first subsequent detectable gaze position. This produced a continuous gaze-coordinate sequence for subsequent spatial mapping.

The identity of the slide was estimated independently of the gaze stream using image matching within-slide AOIs. The video was sampled at 0.5-s intervals for slide-state identification, while the video frame rate was used only to retrieve the frame nearest each sampling time. For each sampled crop, image similarity was evaluated using a multi-region distance,
\begin{equation}
d =
0.45d_{\mathrm{full}}
+0.35d_{\mathrm{part}}
+0.20d_{\mathrm{hist}},
\label{eq:slide_matching}
\end{equation}
where $d_{\mathrm{full}}$, $d_{\mathrm{part}}$, and
$d_{\mathrm{hist}}$ denote the full-region, partial-region, and
histogram-based image distances, respectively.
The weights were heuristically selected to prioritize overall visual similarity while incorporating regional and histogram-based information. The coefficients were normalized to sum to one to maintain a consistent scale for the combined distance measure.

A sampled crop was treated as a new keyframe only when its distance from
the previous key crop was $d \geq 1.2$. Each detected keyframe was then
compared with the slide references in the shared library and assigned to
the reference with the minimum matching distance. A new slide assignment
was accepted only when the best matching distance was $d \leq 18$;
otherwise, the previously assigned \texttt{slide\_id} was retained.
Because no previous slide state was available for the first keyframe, it
was always assigned to its best-matching reference. To reduce rapid
switching between visually similar references, a stickiness margin of
$1.5$ was applied, such that the current slide assignment was retained
unless an alternative reference provided a sufficiently lower matching
distance. The resulting assignments divided each recording into contiguous slide
windows and can be expressed as
\begin{equation}
T(t)\rightarrow S_m,
\end{equation}
where $T(t)$ denotes recording time and $S_m$ denotes the slide assigned
to that interval.

The continuous gaze sequence was then synchronized with the slide-window sequence by timestamp. For each gaze sample, the active slide $S_m$ was first identified, after which the gaze coordinate was tested against the AOI polygons defined for that slide using a point-in-polygon operation. The resulting observation was represented as
\begin{equation}
X(t)=(S_m,E_{mn}),
\end{equation}
where $S_m$ identifies the current slide and $E_{mn}$ identifies the corresponding AOI. Gaze samples falling outside all explicitly labeled AOIs were assigned to the slide-specific unrelated region.

This process produced a temporally ordered multilevel representation of learner interaction that linked frame-level gaze position, within-slide AOI attention, and cross-slide navigation. Consecutive samples assigned to the same slide-element state were additionally collapsed into continuous dwell segments for duration-based analyses. The resulting representation supported subsequent computation of gaze-time allocation, AOI coverage, switching and transition measures, entropy metrics, revisitation, slide-navigation trajectories, and multilevel graph analyses.

\subsection{Eye-Tracking Metrics}

We computed four families of metrics, including AOI-based, entropy-based, slide-level navigation, and multilevel graph metrics, to characterize gaze allocation and visual attention distribution, and learner navigation patterns. Within-slide metrics were computed for each participant $p$ and slide $m$.

\subsubsection{AOI-Based Metrics}

AOIs were defined using the shared slide-element library. Let $N_m$ denote the number of defined instructional and interface AOIs on slide $m$, excluding the catch-all unrelated state $E_0$. For each slide-specific AOI $E_n$, labeled gaze time was calculated by summing the sampling intervals assigned to that element,
\begin{equation}
G_{p,m,n}
=
\sum_{t:X(t)=E_n}\Delta t ,
\label{eq:gaze_time}
\end{equation}
where $\Delta t$ is the duration represented by each valid labelled gaze sample. Because slides differed in total viewing time, we also calculated the gaze-time ratio
\begin{equation}
GR_{p,m,n}
=
\frac{G_{p,m,n}}{T_{p,m}},
\label{eq:gaze_ratio}
\end{equation}
where $T_{p,m}$ is the total time participant $p$ spent viewing slide $m$. When a participant visited the same slide more than once, $T_{p,m}$
and the corresponding slide-level gaze metrics were calculated by
pooling observations across all visits to that slide. Transition counts
were accumulated across visits.

AOI coverage was calculated as
\begin{equation}
C_{p,m}
=
\frac{N_{\mathrm{visited},p,m}}{N_m},
\label{eq:coverage}
\end{equation}
where $N_{\mathrm{visited},p,m}$ is the number of distinct defined AOIs viewed by participant $p$ on slide $m$. The unrelated state $E_0$ was excluded from both the numerator and denominator of the coverage measure.

The gaze switch was represented using a directed transition matrix
$\mathbf{P}^{(p,m)}=[p_{ij}]$, where
\begin{equation}
p_{ij}
=
\frac{N(E_i \rightarrow E_j)}
{\sum_{k\neq i} N(E_i \rightarrow E_k)},
\qquad i\neq j .
\label{eq:transition_probability}
\end{equation}
Thus, $p_{ij}$ represents the conditional probability of transitioning to AOI $E_j$ given a switch away from $E_i$. Transitions were counted only between adjacent temporally ordered samples assigned to the same slide window. Consecutive samples assigned to the same AOI were not treated as transitions, and transitions were not created across slide-window boundaries.

The slide-level switching rate was defined as
\begin{equation}
SR_{p,m}
=
\frac{N_{\mathrm{AOI\ transitions},p,m}}{T_{p,m}} .
\label{eq:switching_rate}
\end{equation}
For participant-level summaries, an overall switching rate was calculated as the total number of valid within-slide AOI transitions across the recording divided by the participant's total recording duration.

We also summarized dwell events at the AOI-category level. Because the supplied dwell-event files contained AOI-category labels but not individual element identifiers, sustained-dwell durations were aggregated by category. Category-level dwell proportions were calculated relative to each participant's total sustained-dwell duration and were used to summarize attention allocation across categories such as assessment, navigation, instruction, and image elements. These sustained-dwell measures were analyzed separately from the element-level labelled gaze-time measures described above.

\subsubsection{Entropy-Based Metrics}

Entropy metrics were used to characterize the distribution and organization of gaze within slides. Let
\begin{equation}
K_m=N_m+1
\end{equation}
denote the total number of gaze states defined for slide $m$, consisting of the $N_m$ defined AOIs and the unrelated state $E_0$. For participant $p$ on slide $m$, the proportion of labelled gaze time assigned to state $i$ was
\begin{equation}
\pi_{p,m,i}
=
\frac{G_{p,m,i}}
{\sum_{j=1}^{K_m}G_{p,m,j}} .
\label{eq:state_probability}
\end{equation}
States receiving no labeled gaze contributed probability zero.

Stationary gaze entropy describes how broadly gaze was distributed across the available states:
\begin{equation}
H_S
=
-\sum_{i=1}^{K_m}
\pi_i\log_2\pi_i,
\qquad
H_S^{*}
=
\frac{H_S}{\log_2 K_m}.
\label{eq:stationary_entropy}
\end{equation}
Values of $H_S^{*}$ closer to zero indicate that gaze was concentrated on a smaller subset of available elements, whereas values closer to one indicate that labelled gaze time was distributed more broadly across the available state space.

Transition entropy was calculated from the switch-only transition matrix
in Eq.~\eqref{eq:transition_probability}:
\begin{equation}
H_T =
-\sum_i \pi_i
\sum_j p_{ij}\log_2 p_{ij},
\qquad
H_T^{*}
=
\frac{H_T}{\log_2(K_m-1)}.
\label{eq:transition_entropy}
\end{equation}
Since self-transitions were excluded, each gaze state could transition
to at most $K_m-1$ alternative states. $\log_2(K_m-1)$
represents the theoretical maximum entropy of the switch-only transition
distribution. Lower values of $H_T^{*}$ indicate that, conditional on
switching, gaze transitions were concentrated among a smaller set of
destinations, whereas higher values indicate more distributed switching
patterns.

Transition entropy was calculated only for participant--slide observations containing at least 10 valid AOI transitions. Observations below this threshold were treated as insufficient for estimating transition entropy. For participant-level descriptive summaries, $H_S^{*}$ and AOI coverage were summarized using the median across all observed slides, whereas $H_T^{*}$ was summarized using the median across slides meeting the minimum-transition criterion. Entropy measures were interpreted descriptively as indicators of visual distribution and transition organization.

\subsubsection{Navigation Metrics}

Cross-slide navigation was derived from each participant's chronological sequence of recorded slide states. Adjacent windows assigned to the same slide were treated as a single continuous visit. For each participant, we recorded total viewing time per slide, total slide visits, the number of distinct slides visited, and the number of revisits. A revisit was defined as entry into a slide that had appeared previously in the same participant's recorded sequence.

Directed slide transitions were defined between consecutive, nonidentical slide visits. For each participant, we constructed a directed slide-transition count matrix $\mathbf{N}^{(p)}=[N_{ij}^{(p)}]$, where $N_{ij}^{(p)}$ denotes the number of observed transitions from slide $i$ to slide $j$. To characterize how broadly each transition was shared across participants, participant prevalence was calculated as
\begin{equation}
PV_{ij}
=
\sum_{r=1}^{P}
\mathbb{I}
\left[
N_{ij}^{(r)}>0
\right],
\label{eq:transition_prevalence}
\end{equation}
where $P$ is the number of participant records and $PV_{ij}$ represents the number of participants exhibiting the transition from slide $i$ to slide $j$. Participant prevalence was used to distinguish navigation pathways.

We additionally identified immediate return patterns of the form
\begin{equation}
S_i \rightarrow S_j \rightarrow S_i,
\qquad i\neq j,
\label{eq:aba_return}
\end{equation}
within each participant's chronological sequence. These $A\!-\!B\!-\!A$ patterns capture localized back-and-forth navigation and were calculated directly from individual sequences.

\subsubsection{Multilevel Graph Metrics}

Three analytically distinct graphs were constructed and linked in a multilevel representation: a participant-record graph, a slide-navigation graph, and an AOI-transition graph. At the AOI level, nodes represented slide-specific AOI elements and directed edges represented observed within-slide changes between elements. The detailed visual encodings are mapped in Table~\ref{tab:graph_encodings}. At the slide level, nodes represented recorded slide states and directed edges represented adjacent slide transitions. At the participant-record level, nodes represented individual recordings and undirected edges represented similarity in their overall distributions of slide-to-slide transitions.

\begin{table}[t]
\centering
\caption{Quantitative visual encodings used in the multilevel graph.}
\label{tab:graph_encodings}
\scriptsize
\setlength{\tabcolsep}{6pt}
\renewcommand{\arraystretch}{1.08}

\begin{tabular}{lll}
\hline
\textbf{Level} & \textbf{Encoding} & \textbf{Meaning} \\
\hline
Participant-record & Edge width & Navigation similarity $S_{pq}$ \\
Slide & Node size & Mean participant-normalized viewing-time\\
Slide & Arrow width & Participant prevalence $PV_{ij}$ \\
AOI & Node size & Mean participant-normalized labelled gaze time\\
AOI & Node color & AOI category \\
AOI & Arrow width & Participant prevalence \\
Cross-level & Dashed line & Membership/visitation only \\
\hline
\end{tabular}
\end{table}

For each participant $p$, the directed transition counts were normalized
over all observed slide-transition pairs
\begin{equation}
q_{ij}^{(p)}
=
\frac{N_{ij}^{(p)}}
{\sum_{a,b}N_{ab}^{(p)}}.
\label{eq:navigation_distribution}
\end{equation}
Pairwise navigation similarity between participant records $p$ and $q$ was calculated as
\begin{equation}
S_{pq}
=
1-
d_{\mathrm{JS}}
\left(
\mathbf{q}^{(p)},
\mathbf{q}^{(q)}
\right),
\label{eq:js_similarity}
\end{equation}
where $d_{\mathrm{JS}}$ is the base-2 Jensen--Shannon distance\cite{lin1991}. Larger values indicate more similar distributions of directed slide transitions. This measure compares first-order transition distributions and therefore does not represent similarity in complete chronological trajectories or viewing durations.

At the slide and AOI levels, graph summaries included the number of distinct incoming and outgoing neighbors and the corresponding incoming and outgoing transition counts. Reciprocal connections were also identified to describe pairs of states connected in both directions. These measures were used descriptively to characterize highly connected slide states and recurring gaze-switching relationships. Reciprocal edges were not interpreted as evidence of an immediate return unless the corresponding $A\!-\!B\!-\!A$ sequence was observed within an individual participant record.

\subsubsection{AOI Category Network Analysis}
To summarize gaze behavior across element categories, AOIs at the slide level were pooled by their predefined AOI type. Node size in the category-level network represented the mean proportion of sustained living duration at the participant levelategory. Directed edges represented observed changes between AOI categories within the same slide window, and edge width represented participant prevalence of the corresponding transition. To improve readability of the aggregate network, only transition edges observed in at least four of the seven participant recordings were displayed. But all observed transitions were retained when computing transition counts and graph metrics.

\subsection{Qualitative Analysis and Synthesis}
Focus group recordings were transcribed and analyzed using content analysis. Transcripts were read in full and segmented into meaning units about participants' interaction with the lesson. The learners' self-reported pros and cons of the lesson regarding the usability were categorized based on the STP heuristics \cite{jahnke2021}.

Quantitative gaze findings and qualitative findings were integrated for analyzing the usability at interface features and instructional activities levels. These were compared with focus-group comments concerning the same features. Convergence and divergence between the behavioral and self-reported evidence were used to inform the usability interpretation.

\section{Results}

\subsection{Learner behaviors derived from multi-level gaze data}

The participant-level gaze metrics (Table~\ref{tab:gaze_metrics}) showed variation in both the frequency and organization of within-slide visual attention. Overall AOI switching rates ranged from 1.81 to 3.48 transitions per second, with P1 showing the highest switching rate and P6 the lowest. Median normalized stationary entropy ($H_S^{*}$) ranged from 0.274 to 0.459 across participants, indicating differences in how broadly gaze was distributed across slide elements. Median normalized transition entropy ($H_T^{*}$) ranged from 0.168 to 0.296 across participants, suggesting that gaze transitions were generally concentrated within a relatively limited set of recurring pathways. Across all participant--slide observations, median AOI coverage was 0.500, meaning that learner looked at about 50\% of the predefined AOIs on each slide (IQR = 0.366--0.750). Median $H_S^{*}$ was 0.411 (IQR = 0.207--0.562), and median $H_T^{*}$ was 0.258
(IQR = 0.119--0.357). These descriptive results indicate that learners differed in the breadth and frequency of their visual exploration, while their gaze transitions among the elements tended to be less dispersed.

\begin{table}[H]
\centering
\caption{Participant-level summary of gaze metrics. 
$SR$ denotes the overall AOI switching rate, $H_S^{*}$ normalized stationary gaze entropy, 
and $H_T^{*}$ normalized gaze transition entropy. Entropy values represent medians across slides.}
\label{tab:gaze_metrics}
\footnotesize
\setlength{\tabcolsep}{8pt}
\renewcommand{\arraystretch}{1.05}

\begin{tabular}{lccc}
\hline
\textbf{Participant} &
\textbf{SR (/s)} &
$\mathbf{H_S^{*}}$ &
$\mathbf{H_T^{*}}$ \\
\hline
P1 & 3.48 & 0.459 & 0.168 \\
P3 & 1.94 & 0.450 & 0.296 \\
P4 & 2.68 & 0.341 & 0.216 \\
P5 & 2.68 & 0.274 & 0.230 \\
P6 & 1.81 & 0.431 & 0.269 \\
P7 & 2.62 & 0.366 & 0.262 \\
P8 & 2.55 & 0.407 & 0.267 \\
\hline
\end{tabular}
\end{table}

\begin{figure}
    \centering
    \includegraphics[width=1.0\linewidth]{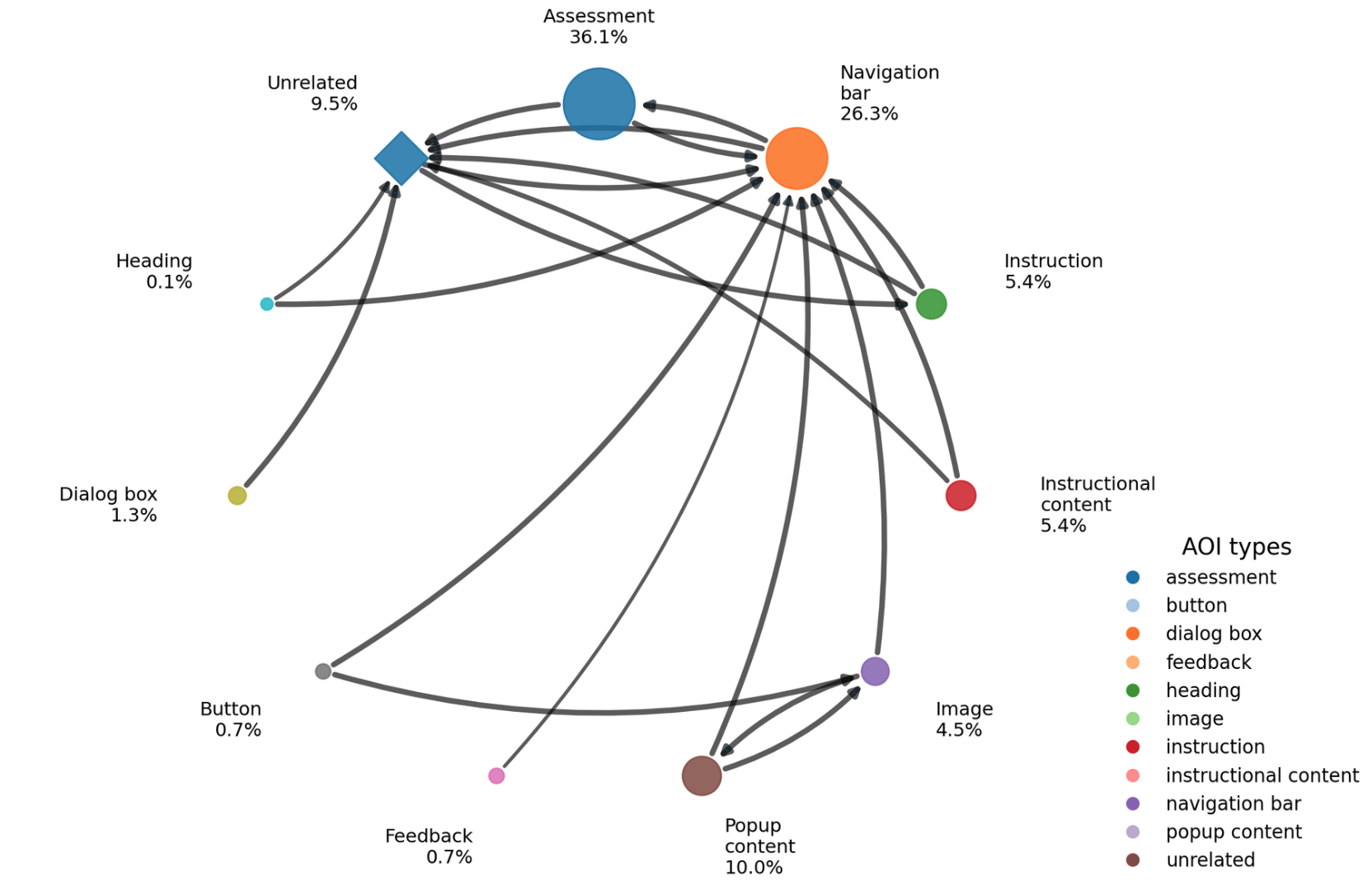}
    \caption{Aggregate gaze allocation and switching AOI categories across seven participants.}
    \label{fig:AOI_category_network}
\end{figure}

The AOI-category network further summarized attention allocation and gaze switching across element types. As shown in Figure~\ref{fig:AOI_category_network}, learners' sustained visual attention was concentrated largely on assessment (36.1\%) and the navigation bar (26.3\%), followed by popup content (10.0\%), unrelated regions (9.5\%), instruction (5.4\%), instructional content (5.4\%), and images (4.5\%). The transition structure showed recurring gaze switching among instructional and interface elements, with the navigation bar functioning as a prominent connection point across several AOI categories. These patterns suggest that learner behavior involved both sustained attention to assessment-related content and repeated switching between content and interface elements during slide-level interaction.

\begin{table*}[t]
\centering
\caption{Aggregate directed transitions among AOI categories across the seven participant recordings. 
Rows represent source AOI categories and columns represent destination AOI categories. 
Each cell reports the total number of adjacent AOI-category label changes, with the number of participants exhibiting the transition in parentheses. 
Self-transitions were excluded.}
\label{tab:aoi_transition_full}
\scriptsize
\setlength{\tabcolsep}{3pt}
\renewcommand{\arraystretch}{1.10}

\resizebox{\textwidth}{!}{
\begin{tabular}{lrrrrrrrrrrr}
\hline
\textbf{Source} &
\textbf{AS} &
\textbf{NB} &
\textbf{IN} &
\textbf{IC} &
\textbf{IM} &
\textbf{PC} &
\textbf{FB} &
\textbf{BT} &
\textbf{DB} &
\textbf{HD} &
\textbf{UR} \\
\hline

AS & -- & 2427 (7) & 103 (7) & 0 & 0 & 0 & 0 & 0 & 0 & 2 (2) & 147 (7) \\

NB & 2388 (7) & -- & 796 (7) & 759 (7) & 772 (7) & 803 (7) &
307 (4) & 73 (7) & 1 (1) & 56 (7) & 2279 (7) \\

IN & 128 (7) & 776 (7) & -- & 3 (3) & 63 (7) & 5 (3) &
0 & 3 (2) & 0 & 11 (6) & 227 (7) \\

IC & 0 & 765 (7) & 0 & -- & 9 (4) & 0 &
0 & 1 (1) & 0 & 1 (1) & 17 (6) \\

IM & 0 & 749 (7) & 66 (7) & 12 (5) & -- & 1246 (7) &
0 & 115 (7) & 0 & 4 (4) & 213 (7) \\

PC & 0 & 817 (7) & 9 (3) & 0 & 1261 (7) & -- &
0 & 24 (7) & 0 & 0 & 32 (7) \\

FB & 0 & 308 (4) & 0 & 0 & 0 & 0 &
-- & 0 & 0 & 0 & 0 \\

BT & 0 & 96 (7) & 3 (3) & 1 (1) & 96 (7) & 21 (7) &
0 & -- & 0 & 0 & 5 (3) \\

DB & 0 & 2 (2) & 0 & 0 & 0 & 0 &
0 & 0 & -- & 0 & 17 (7) \\

HD & 4 (3) & 54 (7) & 16 (6) & 0 & 2 (2) & 0 &
0 & 0 & 0 & -- & 23 (5) \\

UR & 165 (7) & 2237 (7) & 226 (7) & 18 (7) & 213 (7) & 36 (7) &
1 (1) & 3 (3) & 15 (6) & 25 (5) & -- \\

\hline
\end{tabular}
}

\vspace{1mm}
\parbox{\textwidth}{
\scriptsize
\textit{Note.}
AS = assessment; NB = navigation bar; IN = instruction;
IC = instructional content; IM = image; PC = popup content;
FB = feedback; BT = button; DB = dialog box; HD = heading;
UR = unrelated. Values are pooled across slides and participant recordings.
}
\end{table*}

Table~\ref{tab:aoi_transition_full} further quantifies the switching
relationships shown in Figure~\ref{fig:AOI_category_network}. The strongest
bidirectional relationship occurred between assessment and navigation
elements, with 2,427 assessment-to-navigation and 2,388
navigation-to-assessment label changes, both observed in all seven
recordings. Navigation elements were also frequently connected with
instruction, instructional content, images, popup content, and unrelated
regions. A second prominent reciprocal relationship occurred between popup
content and images, with 1,261 popup-to-image and 1,246 image-to-popup
changes, again observed in all seven recordings. These results show that
gaze switching was organized around several recurring instructional--interface
relationships, with the navigation bar serving as a prominent connection
point across multiple AOI categories.

\subsection{Navigation trajectories through the digital learning intervention}

The participant-level navigation metrics (Table~\ref{tab:rq2_navigation_metrics}) showed substantial variation in how learners moved through the intervention. Total slide visits ranged from 49 for P6 to 83 for P8, while unique slides visited were similar across participants, ranging from 37 to 46. Revisitation accounted for 24.5\% to 54.2\% of all slide visits, with P8 showing the highest revisit proportion and P6 the lowest. Immediate $A-B-A$ return patterns also varied, ranging from 6 occurrences for P6 to 22 for P8. These results indicate that differences in navigation were driven less by the number of distinct slides accessed than by how frequently learners returned to previously viewed slides and engaged in localized back-and-forth navigation.

\begin{table}[t]
\centering
\caption{Participant-level summary of slide-navigation metrics. 
Visits indicate total recorded slide visits, unique slides indicate the number of distinct slide states observed, 
revisits indicate returns to previously viewed slides, and $A\!-\!B\!-\!A$ returns indicate immediate return sequences.}
\label{tab:rq2_navigation_metrics}
\footnotesize
\begin{tabular}{lrrrrr}
\hline
\textbf{Participant} &
\textbf{Visits} &
\textbf{Unique} &
\textbf{Revisits} &
\textbf{Revisit} &
\textbf{$A\!-\!B\!-\!A$} \\
&
&
\textbf{Slides} &
&
\textbf{(\%)} &
\textbf{Returns} \\
\hline
P1 & 67 & 46 & 21 & 31.3 & 9 \\
P3 & 71 & 38 & 33 & 46.5 & 19 \\
P4 & 66 & 40 & 26 & 39.4 & 12 \\
P5 & 63 & 38 & 25 & 39.7 & 7 \\
P6 & 49 & 37 & 12 & 24.5 & 6 \\
P7 & 58 & 41 & 17 & 29.3 & 9 \\
P8 & 83 & 38 & 45 & 54.2 & 22 \\
\hline
\end{tabular}
\end{table}

\begin{figure}
    \centering
    \includegraphics[width=0.75\linewidth]{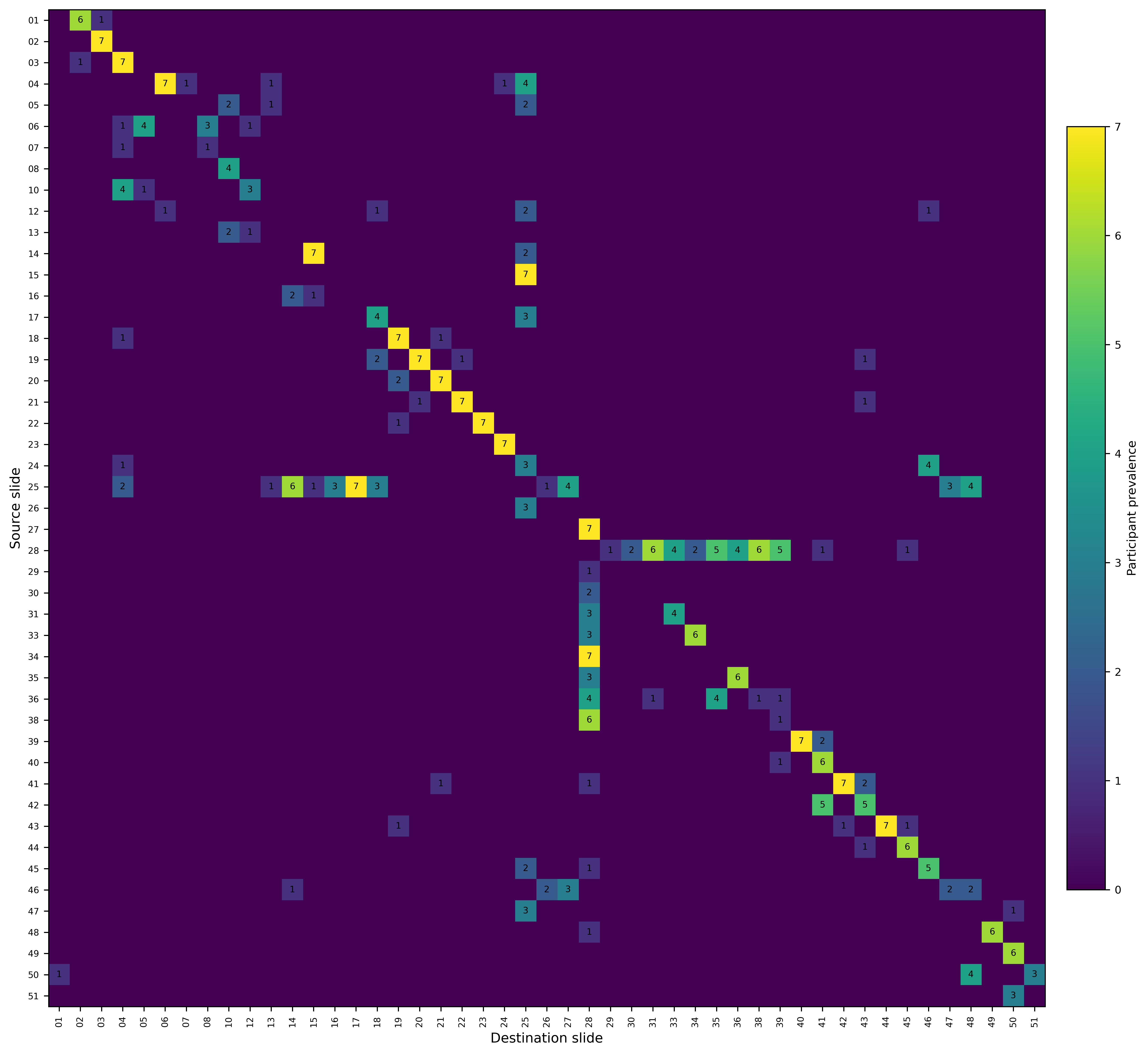}
    \caption{Full slide-navigation transition matrix across seven participants.
Rows indicate source slides and columns indicate destination slides.
Cell values represent participant prevalence for each directed transition;
blank cells indicate unobserved transitions.}
    \label{fig:navigation_transition_matrix_full}
\end{figure}

The navigation transition matrix (Figure~\ref{fig:navigation_transition_matrix_full}) showed both shared navigation pathways and localized branching and return patterns across participants. Forty-seven directed slide transitions were shared byunique slides visited werensitions were observed in all seven recordings, including Slide~2$\rightarrow$3 (slides with instructional materials with an image embedded), Slide~3$\rightarrow$4 (from instructional material to interactive diagram), Slide~14$\rightarrow$15 (drag-and-drop assessment), Slide~18$\rightarrow$19 (from multiple-choice question 1 to its feedback), Slide~19$\rightarrow$20 (from multiple-choice question 1 feedback to multiple-choice question 2), Slide~20$\rightarrow$21  (from multiple-choice question 2 to its feedback), Slide~21$\rightarrow$22 (from multiple-choice question 2 feedback to multiple-choice question 3), Slide~22$\rightarrow$23 (from multiple-choice question 3 to its feedback), and Slide~23$\rightarrow$24 (from multiple-choice question 3 to human typing assessment). In contrast, Slide~28 (interactive diagram) showed greater branching, with shared transitions to interactive-diagram popups, including slides~31, 33, 35, 36, 38, and multiple-choice assessment slide 39, as well as return transitions from Slides~34, 36, and 38 (interactive-diagram popups). These patterns suggest that learners followed partially shared navigation routes while diverging and revisiting particular slide states at specific points in the intervention.

\begin{figure}
    \centering
    \includegraphics[width=1\linewidth]{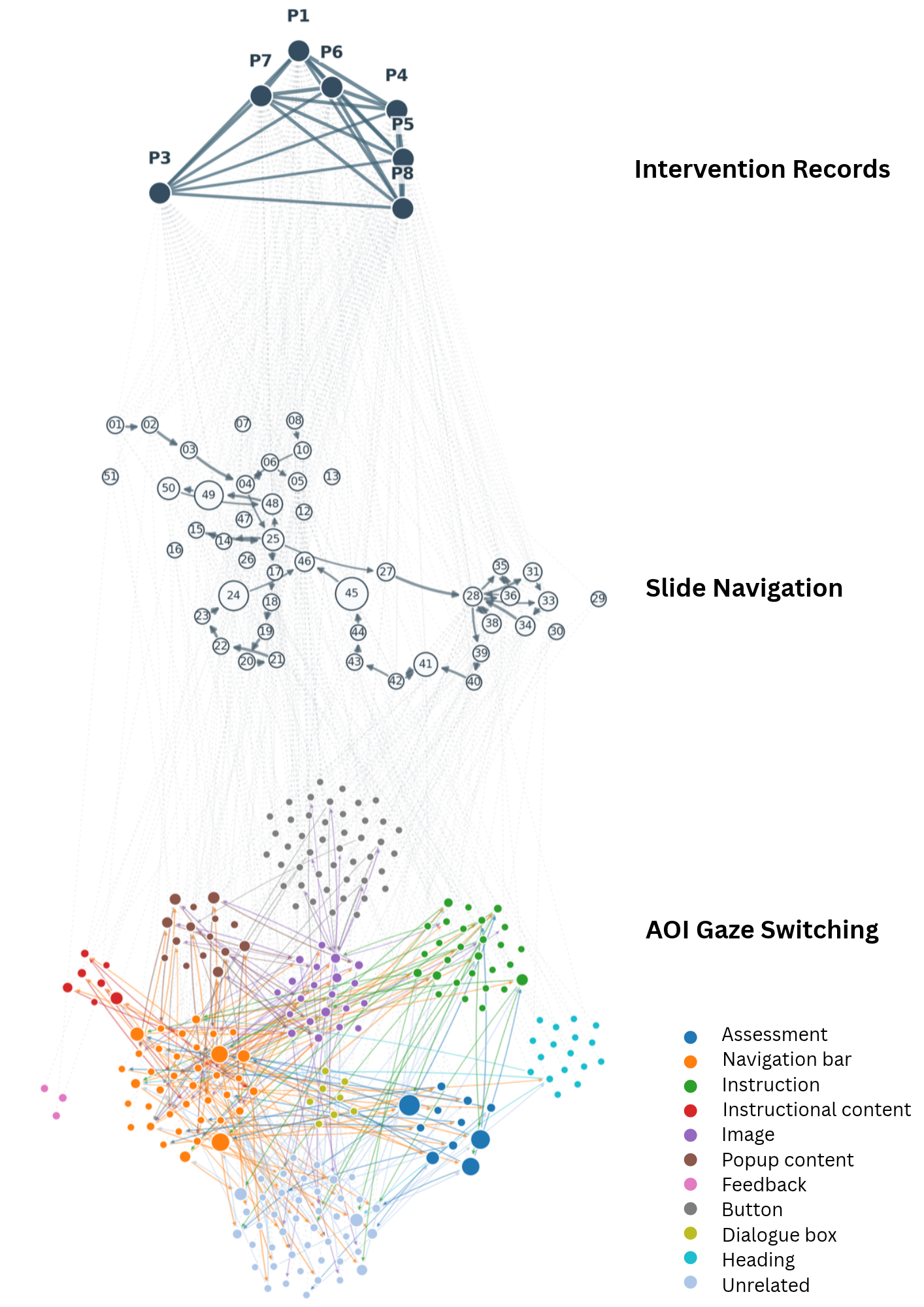}
    \caption{Aggregate three-level network of participant-record similarity, slide navigation, and AOI switching}
    \label{fig:three_level_graph}
\end{figure}

The three-level graph analysis (Figure~\ref{fig:three_level_graph}) revealed related but distinct patterns at the participant-record, slide, and AOI levels. This analysis comprised seven participant-record nodes with 21
pairwise navigation-similarity relationships, 47 slide nodes connected by
127 distinct directed navigation transitions, and 237 slide-specific AOI
nodes connected by 654 directed within-slide gaze transitions. For
visual readability, the aggregate graph displays slide and AOI transitions
observed in at least four of the seven participant records, while all
observed edges were retained for metric computation. 

\begin{figure}
    \centering
    \includegraphics[width=0.8\linewidth]{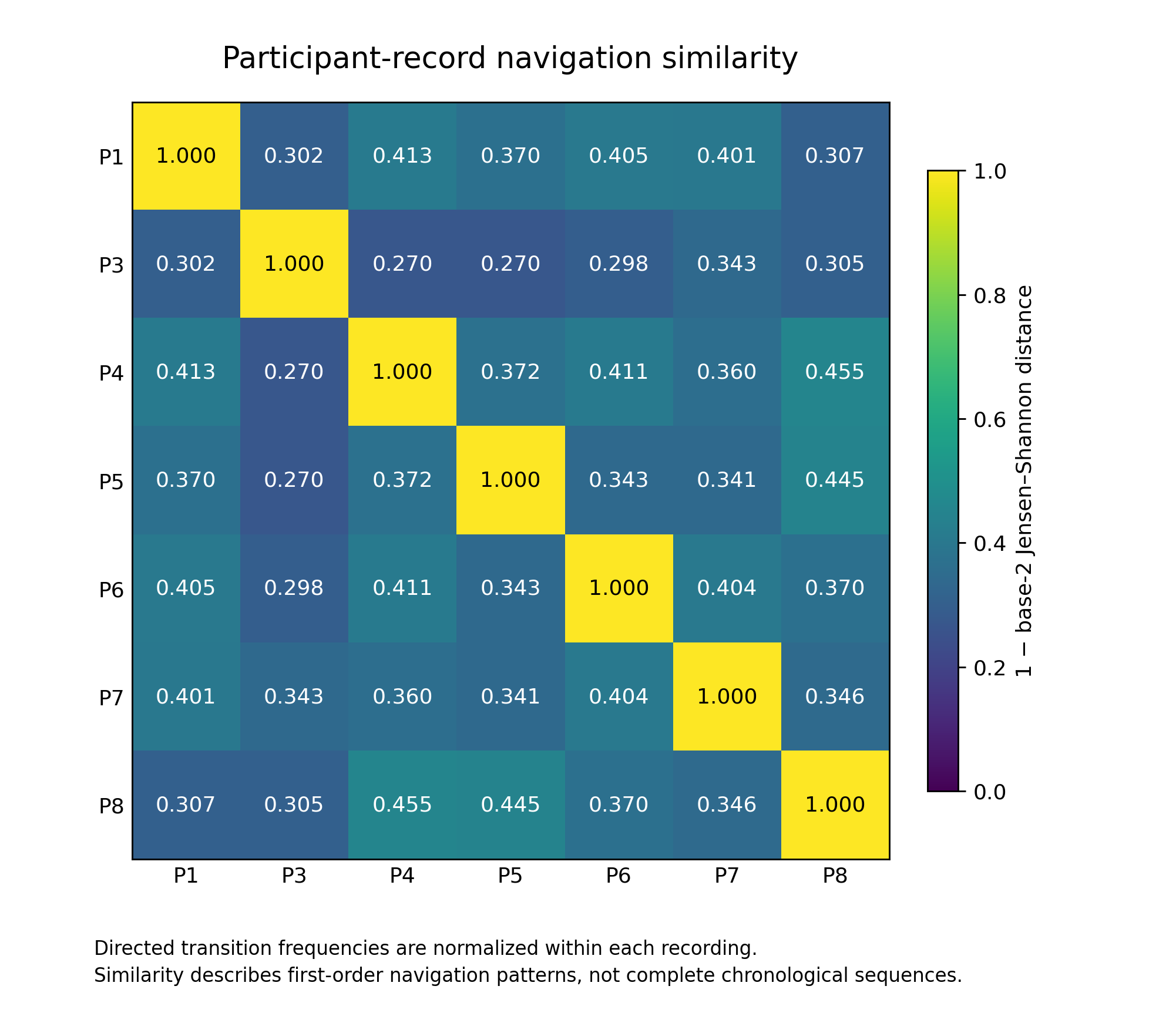}
    \caption{Pairwise similarity of slide-navigation patterns across seven participant recordings}
    \label{fig:participant_similarity_matrix}
\end{figure}

At the participant-record level, the pairwise similarity matrix (Figure~\ref{fig:participant_similarity_matrix}) showed variation in slide-navigation patterns across the seven participant recordings. The highest similarity was observed between P4 and P8 ($S_{pq}=0.455$), followed by P5 and P8 ($S_{pq}=0.445$) and P1 and P4 ($S_{pq}=0.413$). Higher values indicate more similar distributions of directed slide-to-slide transitions, meaning that these participant pairs used similar navigation connections with more comparable relative frequencies.

At the slide level, the seven recordings contained 450 slide-transition
occurrences distributed across 127 distinct directed connections. Slide~28
emerged as a recurrent navigation state: it could be reached from 12
different slides and followed by 11 different slides, and 42 of its visits
were revisits after the slide had already appeared earlier in a
participant's sequence. In addition, 84 immediate
$A\!-\!B\!-\!A$ return patterns were observed across the recordings,
indicating localized back-and-forth movement.

At the AOI level, 21,074 adjacent gaze-label changes formed 654 distinct
directed element-to-element connections. A particularly consistent pattern
occurred on Slide~45 between the assessment and navigation-bar AOIs:
assessment-to-navigation changes occurred 813 times and
navigation-to-assessment changes occurred 796 times, with both directions
observed in all seven participant recordings. These counts represent
changes between adjacent AOI labels.

\subsection{Usability of the digital learning environment}
We examined focus-group responses using a sociotechnical-pedagogical (STP) heuristic \cite{jahnke2021}, which provides heuristics for identifying online learning usability through social, technical, and pedagogical dimensions. Participants' self-reported benefits and limitations regarding this self-guided online intervention's usability were revealed across pedagogical and technological dimensions. We classified relevant excerpts according to seven STP heuristics on the intervention's benefits and limitations, as shown in Table~\ref{tab:usability_findings}.

\begin{table*}[t]
\centering
\caption{Usability Benefits and Limitations Based on the STP Usability Framework}
\label{tab:usability_findings}

\begingroup
\footnotesize
\setlength{\tabcolsep}{5pt}
\renewcommand{\arraystretch}{1.25}

\begin{tabularx}{\textwidth}{
@{}
>{\raggedright\arraybackslash}p{3.2cm}
>{\raggedright\arraybackslash}X
>{\raggedright\arraybackslash}X
@{}
}
\hline
\textbf{STP Heuristic} &
\textbf{Reported Benefits} &
\textbf{Reported Limitations} \\
\hline

\multicolumn{3}{@{}l}{\textbf{Technological Usability}} \\
\hline

Easy to Use
& "The interface was nice like it was just a nice platform."
& --- \\

Page Layout
& "I like the overall look of the lesson."
& "There were, some icons, that were hard to find."
"I have to click all the icons to move forward, and there were, some icons, that were hard to find." "The question mark on the graph was difficult to associate with the diagram."
"Make it (the feedback) colorful a little bit, like green for what you got it right, and red for what you need more…"\\

Navigation
& "I know where to click to move forward and backward"
& "I wanted to go back and read it over again, but then when I tried to go back to the actual question, it didn't let me."
"It would take me back to like my first question. Then I would have to press submit on all the other questions just to get back to where I was."\\

\hline
\multicolumn{3}{@{}l}{\textbf{Pedagogical Usability}} \\
\hline

Diverse Material / Material Quality
& ``I liked how you could have, like, the graphics and then you're able to, like, click the icons and learn, like, more information about it.''
& --- \\

Material Organization / Material Delivery
& "I feel like you could, like, teach it to anybody unfamiliar with the topic." "I think that would be really easy for them to understand."
& ``I found myself moving so fast and not fully consume the content\ldots{} I hope there are things that help me engage, like, engage on the reading" \\

Assessment
& ``I get told what was wrong and what was correct. So that was like, OK, then I can fix my mistakes for future use."
"The tone (of the feedback) was very encouraging, I was like, ok, I knew I did not respond well to the question, but it told me that I answered something right."
& "If you can give an opportunity (for students) to rewrite the open-ended responses after \ldots{} the AI feedback, to see if the students actually learn from the AI."
"Maybe like (adding) an example response that would meet all the criterials."
"if AI created the a graph with the information that we put from those phases and then comparing it to the actual graph that they gave us to learn to see to like visualize."\\

Guidance
& ---
& ``I gave very short response to the question because I did not know what was expected to be put there.''
"I have to click all the icons to move forward, and there were, some icons, that were hard to find."\\

\hline
\end{tabularx}

\vspace{1mm}
\parbox{\textwidth}{
\footnotesize
\textit{Note.} A dash indicates that no distinct corresponding benefit or limitation was identified in the available excerpts.
}

\endgroup
\end{table*}

\subsubsection{Technological Usability}
Participants generally perceived positive learning experiences with the learning intervention, especially with regard to its visuals and ease of use. However, some interactive elements, such as information icons and question mark icons, are difficult to locate and understand with their associated diagram components. Navigation functions have different voices. Some participants said the navigation is intuitive, while other participants reported it as another concern. They understood how to move back and forth, but encountered restrictions when trying to revisit previously completed questions. For example, some had to navigate through multiple screens and resubmit responses to return to their previous positions.

\subsubsection{Pedagogical Usability}
Participants revealed a good impression of the interactive instructional materials. for example, the clickable information icons and visuals are beneficial when exploring the instructional content by encouraging more active engagement. The learning material is friendly for participants with varying levels of prior knowledge. However, some participants reported concerns regarding the learning pacing and processing. One student indicated progressing rapidly through the materials without fully engaging with the content.

Participants valued the benefits of AI-generated feedback because it is personalized and explanatory, referencing their original responses and distinguishing correct information from missing or incorrect ones. The encouraging tone is another plus. However, participants proposed some places to improve, such as including visual representations via color-coded feedback, providing exemplar responses, and offering another chance to revise their answers after receiving feedback. Moreover, the instructional guidance can be improved. Some reported uncertainty about the expected depth and detail of open-ended responses. They indicated that the expectations became clearer only after reviewing the AI-generated feedback.

\section{Discussion}

Through linking within-slide AOI attention, across-slide navigation, and similarity among participant recordings in an AI-supported online biology lesson, and correspond these patterns against learners' accounts of usability, this exploratory study used multilevel gaze dynamic to model learner--technology interaction. Three major findings stand out. Learners who completed the same lesson differed in width, frequency, and sequence in their sampled slide elements (RQ1). Learners generally walked through similar lesson content and followed some common slide-to-slide pathways, but they differed considerably in slide visit times and backtracking behaviors(RQ2). Finally, the gaze and navigation patterns across slides and interface elements often aligned with learners' description  in the focus groups, helping identify specific areas for design review(RQ3).

\subsection{RQ1: How do learner within-slide behaviors differ from multi-level gaze data?}

Participants' gaze reached about half of of the AOIs defined on each slide (median coverage 0.500), and their visual attention tended to be concentrated on some elements (median $H_S^{*}=0.411$). Participants learning behaviors differed broadly by distributing their visual attention across the same interface, with median $H_S^{*}$ values ranging from 0.274 to 0.459 across participants. These findings reveals added value of gaze analytics for understanding learner behavior. Although common-use online behavior metric, such as the trace data can indicate whether learners accessed a slide or performed an action \cite{winter2024}, such data do not fully reflect which instructional or interface elements actually received visual attention\cite{baker2020, sharma2020}. Gaze analytics provides a more fine-grained view of how learners allocate attention and process the information\cite{alemdag2018}.

Switching rate and transition entropy captured different aspects of learner behavior. P1 switched between AOIs most frequently (3.48 transitions/s), but the low transition entropy ($H_T^{*}=0.168$) indicates that these switches tended to occur repeatedly among a limited set of AOI pairs. Across aggregated participant behavior, median low transition entropy (0.258) indicates that gaze switching concentrated within a relatively restricted set of recurring pathways. These findings show that gaze allocation and gaze-transition organization represent complementary aspects of visual behavior\cite{cui2024,laborde2026}.

Sustained attention concentrated on assessment (36.1\%) and the navigation bar (26.3\%), which together accounted for roughly 62\% of sustained dwell. The most consistent AOI-level pattern was reciprocal switching between assessment and navigation-bar elements on Slide~45(the assessment requiring learner typing input), observed in all seven recordings (813 and 796 changes in the two directions). This suggests that participants repeatedly moved between the task and the navigation controls. Gaze analytics can't explain alone the usability issues or learning challenges.

\subsection{RQ2: What navigation trajectories do learners follow
through the digital learning intervention?}

Learners covered similar amounts of the lesson (37--46 unique slides) but moved through it differently where the revisits accounted for 24.5\% to 54.2\% of visits, and immediate $A\!-\!B\!-\!A$ returns ranged from 6 to 22. Thus, only assessing the unique-slide coverage would make learners' lesson use appear more similar than it actually was. This finding supports the value of examining sequential navigation behavior to characterize learning processes in digital environments \cite{huber2024}. Moreover, the $A\!-\!B\!-\!A$ measure captured immediate return behavior, identifying cases in which learners moved from one slide to another and then directly returned to the previous slide. This short back-and-forth navigation can unfold learners' deep online learning engagement, as prior studies suggest that such looping and revisitation patterns can reflect meaningful learning processes, such as reviewing content after assessment or returning to relevant material, and may differentiate higher- and lower-performing learners \cite{fan2021learning,flanagan2022}.

All seven participants followed the question--feedback sequence across Slides~18--24, and 47 directed transitions were shared by at least four learners. Navigation diverged at Slide~28, an interactive diagram with 12 distinct incoming and 11 outgoing connections and 42 revisits, branching to popup slides (31, 33, 35, 36, 38) and an assessment slide (39). Some of this hub-and-spoke movement is built into the design, since popups return to the diagram. The other may be caused by the assessment questions and results.

The highest navigation similarity was observed between P4 and P8
($S_{pq}=0.455$). Each recording contained only about 65 slide transitions, while
127 distinct directed transitions were seen across the full dataset,
so individual transition distributions were relatively sparse. P8, who
had the highest revisit proportion, also appeared in the two most similar
participant pairs, showing that navigation similarity reflects the relative
use of slide-to-slide connections rather than the overall amount of
revisiting. In addition, $S_{pq}$ is based on first-order transition
frequencies and therefore does not capture the complete chronological
sequence of navigation or the time spent on each slide.

Finally, within-slide and across-slide behavior varied independently in this sample. Comparing Tables~\ref{tab:gaze_metrics} and~\ref{tab:rq2_navigation_metrics}, P6 had both the lowest switching rate and the lowest revisit proportion, but P3 combined a low switching rate (1.94/s) with a high revisit proportion (46.5\%), and P1 had the highest switching rate with a moderate revisit proportion (31.3\%). These are descriptive observations from seven recordings, and we claim no association even though they illustrate why within-slide and across-slide measures add information that neither provides alone.

\subsection{RQ3: How can gaze patterns inform understanding of the
usability of the digital learning environment?}

Learners reported positive impressions of the interface, interactive materials, and AI-generated feedback, but also raised concerns about finding interactive elements, returning to earlier responses, and understanding what the open-ended assessments expected. Under the STP framework \cite{jahnke2021}, these map onto technological usability and pedagogical usability, showing that the two are intertwined. Gaze and navigation measures indicated where interaction concentrated and recurred, while focus-group accounts indicated how learners experienced those features. Pairing behavioral traces with learner talk make the findings more validated and builds on eye-tracking evaluations of e-learning usability.

\textit{Assessment.} Assessment elements received the most sustained visual attention, and participants generally valued the AI-generated feedback because it was personalized, explanatory, and encouraging. However, some participants reported that they did not fully understand how much detail their responses needed until they received the feedback. They suggested providing exemplar responses and allowing students to revise their answers after receiving feedback. This suggests that students may need clearer expectations before submitting their responses, not only feedback afterward. The usefulness of AI-generated feedback may therefore depend partly on how the assessment is structured, including the guidance provided before submission and the opportunities given for revision. 

\textit{Navigation.} The navigation bar accounted for the second-largest share of sustained dwell, indicating that learners devoted substantial visual attention to navigation controls. Revisits to previously viewed slides were also common, and all seven participants showed switching between the assessment and navigation-bar AOIs on Slide~45. According to the focus group, participants generally understood how to move forward and backward through the lesson, but some reported difficulty returning to previously completed questions because the system sent them back to the first question and required them to resubmit later responses. It suggests that returning to earlier content was an important part of learners' interaction, but the interface did not always support this behavior efficiently. Design refinements could include preserving previous responses and returning learners to the exact position from which they navigated away.

\textit{Interactive materials.} Participants generally appreciated the clickable visuals and interactive diagrams, but some found certain icons difficult to locate or match to the correct part of the diagram. One participant also said the material moved too quickly to fully engage with the content. The navigation pattern around Slide~28 provides additional context for these comments that learners frequently moved between the main interactive diagram and its related popup screens, making Slide~28 a recurring navigation point. Although this pattern shows that learners interacted with the diagram, repeated movement between the main slide and popup content does not by itself indicate that learners fully processed the intended instructional information. Popup content accounted for 10.0\% of sustained dwell, suggesting that these elements received meaningful visual attention. In all, the behavioral and focus-group findings suggest the value of clearer visual links between icons and their associated content, along with more explicit guidance on what learners should examine in the interactive diagram.

\subsection{Implications}

The study showa that gaze and navigation data can help identify where usability problems may occur, while learners' comments help explain why those problems occur. In the aspect of instructional design, the findings point to four possible refinements that could be tested in future versions of the intervention: (1) allow learners to return to earlier content without losing their responses or previous position, (2) provide clearer expectations and exemplar responses before learners submit their answers, (3) allow learners to revise their responses after receiving feedback, and (4) make the relationship between interactive icons and their associated diagram content more explicit.

For researchers, the findings suggest several approaches for analyzing learner interaction in the eye-tracking research. First, integrate multiple aspects of eye-tracking dynamics to examine learner behaviors for a deeper understanding and more comprehensive capture of learner engagement. Second, consider measures of revisiting and immediate $A\!-\!B\!-\!A$ returns alongside the number of unique slides visited. In this study, participants visited similar numbers of unique slides but differed considerably in how often and how quickly they returned to previous slides. Third, frequently revisited slides and frequently occurring AOI transitions can help identify specific parts of an intervention for closer investigation through interviews or gaze-replay activities. The multilevel graph supports this process by connecting participant-level, slide-level, and AOI-level patterns in a single representation. Together, these analyses reveal differences in learner interaction that aggregate or single-level measures alone cannot capture.

\subsection{Limitations and Future Work}
This study used seven eye-tracking recordings from a single
AI-supported biology lesson, making the findings not generalizable to other
learners, subjects, or learning environments. In addition, gaze coordinates
were extracted from screen-recorded gaze indicators, missing detections were
filled using a temporal hold procedure, and slide identity and AOI membership
were estimated computationally. These processing decisions may affect dwell
and transition measures. Adjacent AOI-label changes therefore should be
interpreted as relative indicators of switching behavior rather than exact
fixations or saccades. Additionally, the behavioral measures also cannot explain learners' intentions or learning outcomes. Although focus-group data provided context for interpreting the gaze patterns, we did not focus much on gaze analysis of individual-level learning outcomes.

Future research should test the framework with larger and more diverse
samples and across different learning environments. It should also examine
the robustness of the findings under alternative gaze-processing choices and
link multilevel gaze and navigation patterns with learning outcomes,
clickstream data, and retrospective gaze-replay interviews. Finally, the
design refinements identified in this study should be implemented and
evaluated to determine whether they improve usability and learning.

\section{Conclusion}
 
This study examined learner–technology interaction in an online learning intervention using a multilevel gaze analytics approach that integrates within-slide visual attention with across-slide navigation trajectories. The findings unveil individual differences in gaze allocation, switching patterns, and navigation behaviors, along with recurring AOI transitions and shared navigation pathways across participants. By connecting AOI-level gaze dynamics, slide-level navigation, and participant-level similarities, this approach provided a temporally structured representation of how learners interacted with instructional content and interface elements throughout the intervention.

Integrating gaze-based findings with focus-group feedback further shows the potential of using multilevel analytics to inform technological and pedagogical usability evaluation. This approach provides complementary information to contextualize reported experiences and identify interface and instructional features that inform further investigation. Overall, this study offers an analytical approach to examining learner–technology interaction beyond isolated gaze measures and individual learning screens, with methodological implications for evaluating and refining extended digital learning environments.

\section{Acknowledgments}
The authors thank the students who participated in this study
and research team members who supported participant recruitment, data collection, and implementation of the online learning intervention.

ChatGPT was used during manuscript preparation to assist with
language refinement. All AI-assisted content was reviewed, revised, and
verified by the authors, who take full responsibility for the final
manuscript.

\bibliographystyle{IEEEtran}
\bibliography{references}

\end{document}